# Size of Memory Objects

R. V. R. Pandya
Department of Mechanical Engineering, University of Puerto Rico at Mayaguez, Puerto Rico, PR 00681-9257

**Abstract**

I consider model for flares proposed by P.-G de Gennes (*PNAS*, **101** (44), 15778-15781, 2004) and suggest a range for amplification factor ( $I$ ) for inhibitory neurons for the time evolution of non-divergent generations of excitatory neurons which eventually die out. The *exact* numerical solution of the model for the suggested range of the factor ( $I$ ) then provides minimum number of neurons describing size ( $M$ ) for memory objects. The obtained size is M = 8 to 12.75 and which is larger than the size $M$ = 2 to 4 as obtained theoretically by de Gennes.

Almost two years ago, de Gennes [1] presented a tentative and schematic picture of the olfaction storage system in early mammalians. Within the framework of the picture, each odor is stored inside a small cluster of neurons which are excited, not all at once, but in steps as the flare of excitation generated at $E_0$ neurons travels on the network of neurons. The flare of excitation, initiated by an odor through sequence of mechanisms in nasal epithelium, olfactory bulb, lateral olfactory tract, piriform cortex and subsequent initial excitations of $E_0$ neurons in storage area, eventually dies out in realistic systems. The neurons affected or excited by a flare caused by a particular odor, before it dies out, are supposed to be storing information about that odor. And the number of excitatory neurons represents the size $M$ of the memory object for the odor. For analyzing flares, de Gennes suggested a model, following an argument given by Abeles [2], for time evolution of population of excitatory neurons having amplification factor ( $E$ ) which are simultaneously suppressed by the inhibitory neurons (typically 20% of the population) having amplification factor ( $I$ ). His theoretical analysis yielded M = 2 to 4.

Here I present an analysis of de Gennes model and show that a range of real values for amplification factor for inhibitory neurons ( $I$ ) can be fixed for a given value of amplification factor for excitatory neurons ( $E$ ) so as to have stable realistic system. Then based on *exact* numerical solutions of the model equations for this range of values, I suggest minimum number of excitatory neurons within a flare initiated by excitation of $E_0$ neuron at time $t = 0$. This minimum number of neurons representing size $M$ of a memory object for an odor turns out to be larger than the number as suggested by de Gennes.

**Analysis of Local Flares**

Consider $E_k$ and $I_k$ as number of excitatory and inhibitory neurons, respectively, at generation $k$ and let $E_0$ and $I_0$ are number of excitatory and inhibitory neurons, respectively, at time $t=0$ when $k=0$. Following de Gennes [1] we can write populations of these neurons at later times $t_0$, $2t_0$, . . ., $kt_0$ as

$$e_k = 0.8\,E\,e_{k-1} - I\,i_{k-1} \quad \text{with} \quad e_k = E_k/E_0, \quad i_k = I_k/E_0 \tag{1}$$

and

$$i_k = 0.2\,E\,e_{k-1}. \tag{2}$$

Here $E$ and $I$ are amplification factors for excitatory and inhibitory neurons, respectively. Substituting Eq. (2) into Eq. (1), we obtain

$$e_k = 0.8\,E\,e_{k-1} - 0.2\,I\,E\,e_{k-2} \tag{3}$$

and for which an exponential solution

$$e_k = u\,x^k \tag{4}$$

yields

$$x^2 - 0.8\,E\,x + 0.2\,E\,I = 0 \tag{5}$$

where $x$ may be real or complex number. It should be noted that when $x$ is real and less than one, Eq. (4) suggests non-divergent solution for generations of excitatory neurons which eventually die out. For a given $E$ and $I$, Eq. (5) has two roots $x_-$ and $x_+$, given by

$$x_- = 0.4\,E - (0.16\,E^2 - 0.2\,E\,I)^{1/2} \tag{6}$$

and

$$x_+ = 0.4\,E + (0.16\,E^2 - 0.2\,E\,I)^{1/2}. \tag{7}$$

Now we discuss different cases which lead us to find range for $I$ for stable realistic system.

(i) **Case I**: The condition for both the roots being real is

$$I \leq 0.8\,E \tag{8}$$

and for $I = 0.8\,E$ both roots are identical and given by

$$x_- = x_+ = 0.4\,E. \tag{9}$$

(ii) **Case II**: When $0.4\,E < 1$ and both the roots are real, the value of $I = I^*$ which makes $x_+ = 1.0$ can be given by

$$I^* = 0.8\,E - \frac{(1 - 0.4E)^2}{0.2E} \tag{10}$$

and for this $I = I^*$

$$x_- = 0.8\,E - 1. \tag{11}$$

So for $I > I^*$, both the roots will be less than one and the exponential solution $e_k = u\, x^k$ suggests that $e_k$ will not diverge. This along with inequality (8) suggests the range for $I$ as

$$I^* < I \leq 0.8\, E \tag{12}$$

for non-divergent solution for generations $e_k$, and for $E = 1.8$ it yields

$$1.2222 < I \leq 1.44\,. \tag{13}$$

Now for this range of $I$, model Eqs. (1) and (2) are solved numerically for different values for $I$ and the number of excitatory neurons $N_j$ up to generation $j$, i.e.

$$N_j = \sum_{k=0}^{j} e_k \tag{14}$$

is plotted against $j$ in Fig. 1 for initial conditions $e_0 = 1$, $i_0 = 0.25$ representing existence of 20% of inhibitory neurons at $t = 0$. Similar results are also plotted in Fig. 2 but for initial conditions $e_0 = 1$, $i_0 = 0$ representing absence of inhibitory neurons at the initial time $t = 0$. It should be noted that

$$\lim_{j \to \infty} N_j\, E_0 = E_0 \sum_{k=0}^{\infty} e_k = M, \tag{15}$$

which represents the total number of excitatory neurons during the flare which is initiated by the excitation of $E_0$ neurons at time $t = 0$. Within the framework of de Gennes [1], this $M$ represents size of a memory object when $E_0 = 1$. The Fig. 1 suggests that minimum value for $\lim_{j \to \infty} N_j \cong 8$ occurs for $I = 1.44$ and thus for $E_0 = 1$ minimum value for $M \cong 8$. I should mention that the computation for $I > 1.44$, e.g. $I = 1.441$, produces negative values for $e_k$ at later stage of time and thus 1.44 is the higher limiting value for $I$ for realistic situation. Further, results presented in Fig. 2 suggest minimum values for $M$ as 12.75 when $E_0 = 1$. This increase in $M$, in case of Fig. 2, is due to the absence of inhibitory neurons at time $t = 0$. These predicted values of 8 and 12.75 for $M$, when $E_0 = 1$, are comparatively higher than the values ($M = 2$ to $4$) as suggested by de Gennes theoretically. I now briefly discuss the reason for this difference.

Now, following de Gennes [1], when $E = 1.8$ and $I = 1.44$, Eq. (9) suggests $x_- = x_+ = 0.72$ and substitution of exponential solution given by Eq. (4) into Eq. (15) along with $u = 1$ and $E_0 = 1$ yields theoretical value for $M$ as

$$M = E_0 \sum_{k=0}^{\infty} e_k = \frac{1}{1 - x_+} = 3.57. \tag{16}$$

In Fig. 3, theoretical Eq. (4) and numerical solutions for $e_k$ vs. $k$ are shown. From this figure, it is clear that *exact* numerical solutions of Eqs. (1) and (2), indicated by N1 and N2 in Fig. 3, differ from the assumed exponential solution. There is an initial increase in $e_k$ for the exact numerical solutions and which is not captured by the assumed exponential solution, though the exponential solution fixes properly the upper limit for $I$. This initial increase in the number of excitatory neurons contributes additionally and increases the value of $M$ to 8 or 12.75, depending on the presence and absence of inhibitory neurons at time $t = 0$.

In conclusion, the minimum size of memory object for an odor in olfaction storage system is 8 or 12.75 neurons depending on the initial presence and absence of 20% of inhibitory neurons when the amplification factor $E = 1.8$.

**References:**

1. de Gennes, P. G. (2004) *Proc. Natl. Acad. Sci. USA* **101** (44), 15778 – 15781.
2. Abeles, M. (1991) *Corticonics* (Cambridge University Press, Cambridge, U.K.).

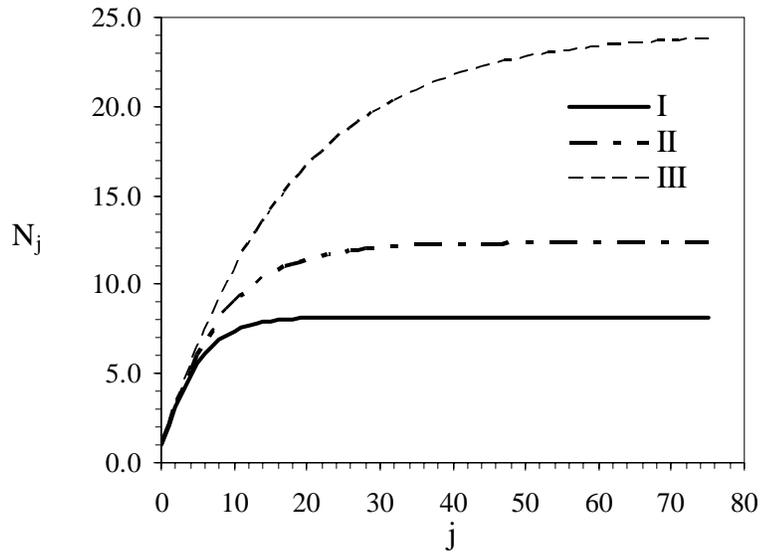

**Fig. 1.** $N_j$ vs. $j$ obtained from numerical solution of Eqs. (1) and (2) for different values for $I$ with $e_0 = 1$ and $i_0 = 0.25$, curve I: $I = 1.44$, curve II: $I = 1.37$, curve III: $I = 1.3$.

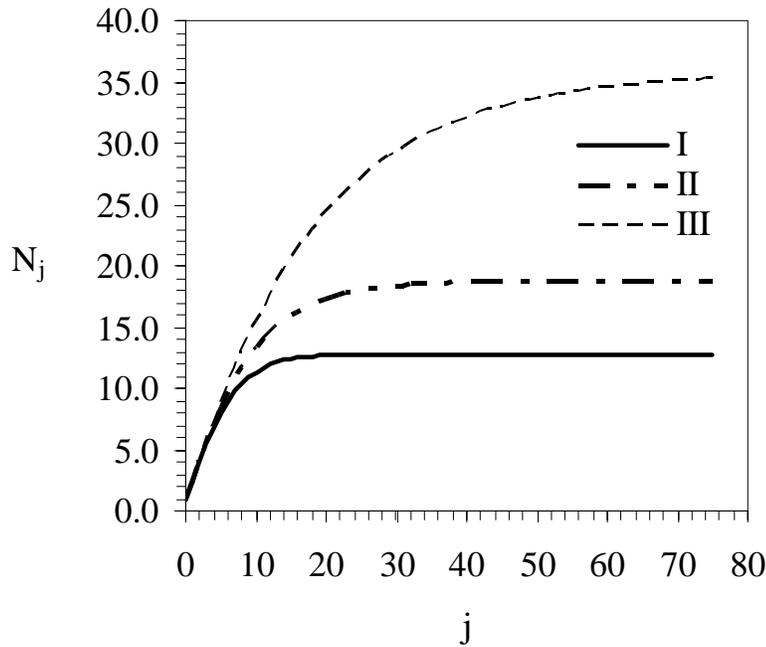

**Fig. 2.** $N_j$ vs. $j$ obtained from numerical solution of Eqs. (1) and (2) for different values for $I$ with $e_0 = 1$ and $i_0 = 0$, curve I: $I = 1.44$, curve II: $I = 1.37$, curve III: $I = 1.3$.

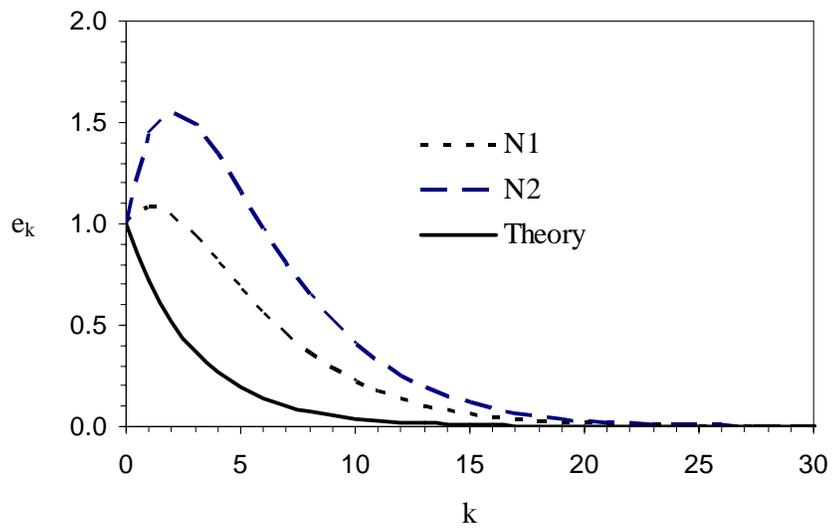

**Fig. 3.** Numerical and theoretical values for $e_k$ vs. $k$. N1, N2 curves are for numerical solutions for $i_0=0.25$ and $i_0=0$, respectively. Theory curve: $e_k = u\, x^k$ with $u = 1$, $x = 0.72$.

-----------------------o0o-----------------------